\documentclass[12pt]{article}
\usepackage[all]{xy}
\usepackage{epic}
\usepackage{amsmath}[1996/11/01]
\usepackage{amssymb,amsthm,amsfonts,latexsym,epsfig,graphics}
 \def\beql#1#2\eeql{\begin{equation}\label{#1}#2\end{equation}}

\usepackage{graphicx}

\newcommand*{\Scale}[2][4]{\scalebox{#1}{\ensuremath{#2}}}

\advance \textheight by -1 \topmargin
\advance \textheight by -1 \headheight
\advance \textheight by -1 \headsep
\advance \textheight by -2 \footskip
\vsize = \textheight
\hsize = \textwidth
\DeclareMathOperator{\depth}{\frak{dp}}
\DeclareMathOperator{\E}{\frak E}
\DeclareMathOperator{\rk}{rk}

\DeclareMathOperator{\Type}{type}

\DeclareMathOperator{\GL}{GL}

\DeclareMathOperator{\Sym}{\mathcal S}
\DeclareMathOperator{\Stab}{Stab}

\DeclareMathOperator{\id}{id}

\newtheorem{theorem}{Theorem}[section]

\newtheorem{fact}[theorem]{Fact}

\newtheorem{prop}[theorem]{Proposition}

\newtheorem{example}[theorem]{Example}
\newtheorem{corollary}[theorem]{Corollary}
\newtheorem{cor}[theorem]{Corollary}

\newcommand{\bew}{\noindent\underline{Proof.}\ }
\newtheorem{rem}[theorem]{Remark}
\newtheorem{remark}[theorem]{Remark}
\newtheorem{lemma}[theorem]{Lemma}

\newtheorem{definition}[theorem]{Definition}
\newtheorem{defn}[theorem]{Definition}

\newcommand{\disj}{\stackrel{.}{\cup}}

\newcommand{\N}{{\mathbb{N}}}

\newcommand{\eb}{\phantom{zzz}\hfill{$\square $}\smallskip}

\newcommand{\cC}{{\mathcal C}}
\newcommand{\cF}{{\mathcal F}}
\newcommand{\cG}{{\mathcal G}}

\title{Network coding and spherical buildings}
\author{ Dirk Liebhold, Gabriele Nebe and Angeles Vazquez-Castro} 
 
\begin{document}
\maketitle
\textsc{
Lehrstuhl D f\"ur Mathematik, RWTH Aachen University,
52056 Aachen, Germany}

\emph{E-mail address}{:\;\;}
\texttt{dirk.liebhold@rwth-aachen.de,}
\texttt{nebe@math.rwth-aachen.de}
\medskip 

\bigskip

\textsc{
Universitat Autonoma de Barcelona}

\emph{E-mail address}{:\;\;}\texttt{angeles.vazquez@uab.es}
\medskip

{\sc Abstract.} 
We develop a network coding technique based on flags of subspaces and 
a corresponding network channel model.
To define error correcting codes we introduce a new distance on the
flag variety, the Grassmann distance on flags and compare it 
to the commonly used gallery distance for full flags. 



\bigskip

\section{Introduction}
\label{intro}

To transmit information over network channels, the 
currently used method 
consists of routing, i.e. simply forwarding the packets through each node. 
Network coding assumes that the 
packets that are sent through the network are 
elements of a vector space and the nodes in the network 
forward linear combinations of the received packets with
randomly selected coefficients. 
It is well known (see \cite{Ahlswede}) 
that network coding allows multicast capacity achievability, which is not possible with packet forwarding only. 
When errors exist, the design of error correction codes for
network-coded communication substantially differs 
from bit-level coding design. This is not only
because of the richer algebraic structure 
but also because networking protocols exist between the physical transmission 
channel and the packet-level communication.
Such protocols motivate packet-level error models. 
%A relevant packet-level error model for network communication is the so called packet erasure model.

In the seminal work \cite{KK} a novel solution to the problem of error and erasure correction 
is tackled over a linear network-coded packet flow. The randomly selected coding coefficients are assumed unknown (i.e. incoherent transmission) and a novel framework of subspace coding is proposed. 
If the packets are elements of the vector space $V$ then
the solution by K\"otter and Kschischang 
stems from the observation that information that is preserved after being
 linearly transformed by the network is the subspace generated by the input vectors,
 which is an element of  the Grassmannian space ${\mathcal G}(V) $, the set of all  subspaces of $V$. 

In this work, we extend the applicability of such framework under the assumption that in-network nodes can keep track of packet sequence numbering,
 as it is the case on the Internet. Under such assumption, 
we propose encoding information over flags,
i.e. chains of subspaces of $V$,
 that are network coded by the in-network
 nodes with the stabilizers of the flags as they traverse the network \cite{A-ITW}.
The set of all flags in $V$ forms a simplicial 
complex, known as the spherical building of the general linear group of $V$. 
We modify the well known geometry of this spherical building 
to develop a minimum distance decoding scheme in the new geometry. 
As the geometry of spherical buildings is governed by 
the symmetric group $\Sym _n$, 
Section 2 is mostly devoted to summarize  the relevant facts 
about symmetric groups. In Section 3 we introduce the basics of 
the flag variety of $V$ and the associated spherical building. 
The major goal of the next section is to 
define a new distance on flags, the 
Grassmann distance (see Definition \ref{Grassdist}),
which is more appropriate 
to measure errors and erasures in the transmission of flags through 
the network. 
The Grassmann distance seems to be much easier to
compute than the commonly used gallery distance.
In the last section, Section 5, 
we present a model for network transmission including errors and erasures 
which allows for the derivation of conditions for code constructions based
on the Grassmann distance on flags.
To set up 
a benchmark for comparing new flag codes to the classical 
situation of subspace codes we 
transfer and generalize the rank metric codes from \cite{SKK} to our situation. 
Other examples for flag codes  of smaller minimum distance are 
given which allow for easy decoding.

\section{Symmetric groups}

The symmetric group $\Sym _n$ is the group of all bijective mappings 
from $\{ 1,\ldots , n \}$ to itself. 
As this group will govern the geometry of the flag variety 
we collect some relevant facts about symmetric groups in this section. 
\subsection{The length and the depth of a permutation}

\begin{definition} \label{lengthdef}
(see for instance \cite[Section 1.6]{Humphreys}) \\
The {\bf length} of a permutation 
$\pi \in \Sym _n$  is 
$$
\ell (\pi ) := \sum _{i=1}^{n} 
|\{ k \in \{ 1,\ldots , i\} \mid \pi(k) > \pi(i) \} | .
$$
\end{definition}

Then the identity  is the unique element of $\Sym _n$ having length 0 and 
the elements $\pi \in \Sym _n$ of length 1 are exactly the transpositions 
$t_i= (i,i+1) (1\leq i \leq n-1) $ interchanging $i$ and $i+1$ and leaving
the other points fixed. 
It is well known that any element of $\Sym _n$ is a product of these $t_i$. 
The length of $\pi \in \Sym _n$ is 
the number of factors in such a minimal expression of $\pi $ as a 
product of the $t_i$ (\cite[Section 1.7]{Humphreys}). 
As all the $t_i$ have order 2, this shows that $\ell (\pi ) = \ell (\pi ^{-1} )$.

\begin{remark}\label{longest}
 There is a unique {\bf longest element} 
$w_0 \in \Sym _n$ with maximal $\ell (w_0) $. 
This element is $w_0 = (1,n)(2,n-1)(3,n-2) \cdots $ and has length
$$  \frac{n(n-1)}{2} = {{n}\choose{2}} = | \{ (i,j) \in \{1,\ldots , n\} ^2 \mid i< j \} | $$
\end{remark}

The most commonly used distance on the flag variety is the
gallery distance (see Definition \ref{galery}). 
This distance is defined using the length 
of a permutation. 
For our purposes it seems to be more appropriate to work with 
the Grassmann distance on flags defined in Definition \ref{Grassdist} 
below. Here we replace the length function by a slightly 
different function called the depth function (see \cite[Theorem 1.1]{Petersen}).

\begin{definition}\label{depth} 
For $\pi \in \Sym _n$  we define
$$ \depth (\pi ) := \sum _{i=1}^{n-1} |\{ k\in \{ 1,\ldots , i \} \mid \pi (k) > i \} |  .$$
Then $\depth : \Sym _n \to \N_0$ is called the {\bf depth function}.
\end{definition}

It is easy to see that $\depth (\pi ) = 0$ if and only if 
$\pi = \id $. Also $\depth (\pi ) = 1$ if and only if $\ell (\pi ) = 1$ 
if and only if $\pi = t_i = (i,i+1)$ for some $1\leq i \leq n-1$. 
More generally we get 

\begin{theorem}\label{lesym} 
(see Observation 2.2 in \cite{Petersen})
For any permutation $\pi \in \Sym _n $ we have 
$$\frac{\ell(\pi) + \ell_{tr}(\pi)}{2} \leq \depth(\pi) \leq \ell(\pi)$$
where $\ell_{tr}(\pi)$ is the smallest number of transpositions needed to write $\pi$.
\end{theorem}

\begin{remark} \label{scoopdist} 
\begin{itemize}
\item[(a)] It is easy to see that $$\depth(\pi ) = \sum _{k=1,k<\pi(k)}^n \pi (k) - k .$$
\item[(b)]  For the longest element $w_0$ from Remark \ref{longest} 
we compute   $$\depth(w_0) = \sum _{k=1}^{\lfloor \frac{n}{2} \rfloor} (n-2k+1) = 
\left\{ \begin{array}{ll} (n/2)^2 & n \mbox{ even } \\
(n-1)(n+1)/4 & n \mbox{ odd } \end{array} \right. $$
(see Sequence A002620 in \cite{OEIS})
\item[(c)] The function $$s(\pi) := \depth(\pi ) + \depth(\pi^{-1}) = 
\sum _{k=1}^n |\pi (k) - k | $$ is known as the sum of distances function.
As we will see in Corollary \ref{esym}  $\depth(\pi ) = \depth(\pi ^{-1}) $ 
so $s(\pi ) = 2 \depth(\pi )$. 
\item[(d)] The number of permutations $\pi \in \Sym _n$ such that 
$\depth (\pi ) = k$ is denoted by $T(n,k) $ in the Sequence A062869 		
   in \cite{OEIS}. 
Sequence A062870 in \cite{OEIS} gives 
$$T(n,k_0) = \left\{ \begin{array}{ll} (\frac{n}{2} !)^2 & n \mbox{ even } \\
(\frac{n(n-1)}{2} !)^2 & n \mbox{ odd } \end{array} \right. $$
where 
$k_0 = \depth(w_0) = \max \{ \depth(\pi ) \mid \pi \in \Sym _n \} .$
\end{itemize}
\end{remark}

\subsection{Young subgroups}

\begin{defn}
Let $T' := (k_1,\ldots, k_{m+1}) $ be a sequence of $m+1$ natural numbers 
$k_i \geq 1$  with $\sum _{i=1}^{m+1} k_i = n$.
The {\bf Young subgroup} 
${\mathcal Y}_{T'} \cong \Sym _{k_1} \times \Sym _{k_2} \times \ldots \times \Sym _{k_{m+1}}$
is the stabilizer in $\Sym _n$ 
of the sequence
$$( \{  1,\ldots , k_1 \} , \{ k_1+1,\ldots , k_1+k_2\} , 
\ldots , \{ k_1+\ldots + k_m +1,\ldots , n \} ) .$$
\end{defn} 

Clearly $|{\mathcal Y}_{T'} | = \prod _{i=1}^{m+1} k_i! $. 
For $m=0$ we get ${\mathcal Y}_{T'} = {\mathcal Y}_{(n)} = \Sym _n $. 
Also if $k_i=1$ for all $i$ 
then ${\mathcal Y}_{T'} = {\mathcal Y}_{(1,\ldots , 1 ) } = \{ \id \}$.

It is well known (see for instance \cite{Jones}) 
that any 
double coset ${\mathcal Y}_{T'} \pi {\mathcal Y}_{T'}$ 
contains a unique element of minimal length.
So these double cosets have  canonical representatives
which we collect in the set $\Sigma _{T'}$:

\begin{definition}\label{doublecosetYoung}
Let $\Sigma _{T'} $ denote the set of representatives 
of minimal length such that
$$\Sym _n = \disj _{\pi \in \Sigma _{T'} } {\mathcal Y}_{T'} \pi {\mathcal Y}_{T'} .$$
\end{definition}

\section{Spherical buildings}

This section provides a constructive approach to the relevant 
facts about the spherical building of the general linear group of a
finite dimensional vector space. 
For most of the proofs and more details we refer to the textbooks 
\cite{Abramenko}, \cite{Humphreys}, and \cite{Taylor}.

\subsection{The flag variety}\label{flag} 

Let \emph{$K$} be a field and $V$ an 
$n$-dimensional vector space over $K$. The
general linear group of $V$, $\GL(V)$, is the group of all linear
automorphisms of $V$ (invertible linear maps from $V$ to itself).

A {\bf flag}
is a set of subspaces $\Lambda := \{ W_i \mid 1\leq i \leq m \}$ of $V $
with $$\{ 0 \} <W_1< \ldots < W_m < V .$$
The {\bf type} of $\Lambda $ is the
set of dimensions
$$\Type (\Lambda ):= \{ \dim(W_i) \mid W_i \in \Lambda  \} \subseteq \{ 1,\ldots , n-1 \} .$$
Let
$$\cF (V) := \{ \Lambda \mid \Lambda \mbox{ is a flag  in } V \} $$
denote the set of all flags in $V$ and for $T\subset \{ 1,\ldots, n-1\}$
let
$$\cF _T(V) := \{ \Lambda \in \cF (V) \mid \Type (\Lambda ) = T \} $$
be the set of all flags in $V$ of type $T$.
Note that the intersection of two flags is again a flag.
The empty set is the unique minimal flag, its type is $\emptyset $. 
The second  minimal flags $\{ W_1 \}$ are the
proper subspaces $W_1$ of $V$.
So the Grassmannian of all $k $-dimensional subspaces 
$$\cG _{k } (V) =\{ 0 < W_1 < V \mid \dim (W_1 ) = k \} $$
is in bijection with the set of flags $\cF _{\{ k \} } (V)$ of
type $\{ k \}$. 
A flag is called {\bf full}, if its type is $\{1,\ldots , n-1\}$.
The set of full flags in $V$ is denoted by $\cF _f(V)$.

To construct a set of canonical representatives of the orbits 
of $\GL(V)$ on $\cF (V)$ we choose and fix once and for all 
a full flag 
$$\Delta _0 = \{ V_1 , \ldots , V_{n-1} \} \in \cF _f(V) $$
such that $\dim (V_i) = i$ and call 
$$B:= \{ g\in \GL(V) \mid V_i g = V_i \mbox{ for all } 1\leq i \leq n-1 \} 
= \Stab_{\GL(V)} (\Delta _0) $$ 
the  standard {\bf Borel subgroup}. 
For $T =\{ d_1,\ldots , d_m \} \subseteq \{ 1,\ldots , n-1 \} $ 
define 
$$\Delta _T := \{ V_i \mid i \in T \} \in \cF _T(V) \mbox{ and } 
P_T := \Stab _{\GL(V) } (\Delta _T ) .$$
The groups $P_T$ are called the standard {\bf parabolic subgroups} of 
$\GL(V)$. 

\begin{remark} 
If $T_1\subseteq T_2$, then $P_{T_2} \subseteq P_{T_1} $.
We have $P_{\emptyset} = \GL(V) $ and $P_{\{ 1,\ldots ,n-1\} } = B$.
\end{remark} 

We summarize the situation by listing some important points:

\begin{fact}\label{orbits}
\begin{itemize}
\item[(a)]
The group $\GL (V) $ acts on the set $\cF (V)$.
\item[(b)]
The $\GL(V) $-orbits are separated by the type, so the partition
$$\cF (V) = \bigcup _{T\subseteq \{ 1,\ldots , n-1 \} } \cF _T(V) $$
is a partition of $\cF (V) $ into $\GL(V)$-orbits.
In particular $$\cF_T(V) = \{ \Delta _T g\mid g\in \GL(V) \}. $$
\item[(c)] 
For a given type $T\subseteq \{ 1,\ldots , n-1 \}$, 
the map 
$$
\cF_T(V) \to P_T \backslash \GL(V), 
 \Delta _T g \mapsto  P_Tg  $$
is a bijection between the set of all flags of type $T$ and the 
set 
$$P_T \backslash \GL(V) = \{ P_T g \mid g\in \GL(V) \} $$
of right cosets of $P_T$ in $\GL(V) $.
\end{itemize}
\end{fact}

To define a geometry on the flag variety we want to 
study $\GL(V)$-invariant distance functions on
$\cF_T(V) $.

\begin{rem}\label{distconst} 
Let $T$ be some type and $P_T = \Stab _{\GL(V)} (\Delta _T)$ the
standard parabolic subgroup of $\GL(V)$. If 
$d$ is some $\GL(V)$-invariant function on ${\cF }_T(V)\times {\cF }_T(V)$, (so
$d(\Lambda g, \Lambda ' g) = d(\Lambda,\Lambda ') $ for all $g\in \GL(V)$
$\Lambda , \Lambda '\in {\cF }_T(V)$),
then $d(\Delta_Tg,\Delta_Th) = \overline{d}(hg^{-1}) $ where 
$$\overline{d} (g) 
=d(\Delta _T ,\Delta _T g) \mbox{ for all } g\in \GL(V).$$
Moreover $\overline{d}$ 
is constant on the double coset $P_T gP_T$.
\end{rem}

\bew
As $d$ is $\GL(V)$-invariant we see that 
$$d(\Delta _T g,\Delta _T h) = d(\Delta _T , \Delta _T hg^{-1} )  
= \overline{d} (hg^{-1} ).$$
To see the second assertion 
let $b_1,b_2\in P_T$, $g\in \GL(V)$.
Then
$$\overline{d} (b_1gb_2) = d(\Delta _T , \Delta _T (b_1 g b_2) ) =
 d(\Delta _T b_2^{-1} , \Delta _T b_1  g)
= d(\Delta _T, \Delta _T g ) = \overline{d} (g) .$$
\eb

As different double cosets are disjoint, we obtain a partition
$$\GL(V) = \disj  P_T g P_T $$ of the group $\GL(V)$ into a disjoint union of
double cosets. The number of these double cosets does not depend on the field $K$.
For $T= \{1,\ldots , n-1\}$  this number  is always
$n!$ and there is a canonical bijection between these double
cosets and the group $\Sym _n$ of permutations of $\{1,\ldots, n\}$
where $n=\dim(V)$. Here we embed $\Sym _n$ as the set of permutation matrices into $\GL(V)$. 
This is captured by the Gau\ss-Bruhat decomposition. 
For any type $T$, 
 the $P_T$ double cosets in $\GL (V)$ are in bijection with the 
double cosets of the Young subgroup ${\mathcal Y}_{T'} $ in the 
symmetric group $\Sym _n$.

\begin{theorem}\label{GBgen} (see \cite{Humphreys}, \cite{Jones}) 
$$\GL(V) = \disj _{\pi \in \Sym_n} B \pi B .$$
More generally for a given type
$T =  \{ d_1,\ldots , d_m \} \subseteq \{ 1,\ldots , n-1 \} $ with $0<d_1<\ldots < d_m < n $ 
 we define $T':= (k_1,\ldots , k_{m}, k_{m+1} )$ 
with 
$$k_1:= d_1, k_i = d_i - d_{i-1} \mbox{ for } 2\leq i \leq m \mbox{ and } 
k_{m+1} := n-d_m .$$
Then 
$$ P_T = \disj _{\pi \in {\mathcal Y} _{T'}} B\pi B $$ 
for the Young subgroup ${\mathcal Y}_{T'}$ and 
$$\GL (V) = \disj _{\pi \in \Sigma _{T'}} P_T \pi P_T $$
where $\Sigma _{T'} $ is defined in Definition \ref{doublecosetYoung} 
\end{theorem}

The Gau\ss-Bruhat decomposition  has a very nice property, as described in 
\cite[Theorem 5.10]{Taylor}: 
For each $\pi \in \Sym _n$ there is a subgroup $U_{\pi} \leq B$ such that 
any element in $B\pi B$ has a unique expression as $b\pi u$ with $b\in B$ and $u\in U_{\pi }$. 
If $K$ is a finite field, then the order of $U_{\pi }$ is $|K|^{\ell(\pi )} $ 
where $\ell $ is the length function on $\Sym _n$ (see Definition \ref{lengthdef}).

\subsection{Buildings and apartments} 

To get a more precise model of the geometry of all flags, the 
so called spherical building of the group $\GL(V)$, we 
fix a basis $E:=\{ e_1,\ldots , e_n\} $ of $V$ and put
$$\Delta _0 = \{ V_1,\ldots , V_{n-1} \} 
\mbox{ with } 
V_i = \langle e_1,e_2,\ldots , e_{i} \rangle .$$
We identify $\GL(V) $ with the group of invertible $n\times n$-matrices 
$\GL_n(K)$ using coordinate rows with respect to this basis. 
Then $B = \Stab_{\GL(V)} (\Delta _0)$ is identified with the 
group of all lower triangular matrices in $\GL_n(K)$ 
and the parabolic subgroup $P_T$ with all lower block triangular
matrices 
$$\left( \begin{array}{ccccc} 
A_{11} & 0 & 0 & 0 & 0 \\
A_{21} & A_{22} & 0 & \ldots & \vdots  \\
\vdots & \vdots & \ddots & \vdots & \vdots \\
A_{m1} & \ldots & \ldots & A_{mm} & 0 \\
A_{m+1,1} & \ldots & \ldots & A_{m+1,m} & A_{m+1,m+1} \end{array} \right)  
$$
where 
$
A_{ij} \in K^{k_i \times k_j } \mbox{ and } 
 A_{ii} \in \GL_{k_i} (K) \mbox{ for } 1\leq j\leq i \leq m+1  $
if  $T' = (k_1,\ldots , k_{m+1} ) $.
For a permutation $\pi \in \Sym_n $ we denote by 
$$\Delta _{\pi } := \{ \langle e_{\pi(1)} \rangle , \langle e_{\pi(1)},e_{\pi(2)} \rangle ,
\ldots \langle e_{\pi(1)},e_{\pi(2)},\ldots , e_{{\pi(n-1)}} \rangle \} $$
the full flag constructed by reordering the basis vectors in $E$ according to $\pi$. 
For $\pi \in \Sym_n$ let $\tilde{\pi }\in \GL_n(K)$ denote the corresponding
permutation matrix so that $\Delta _{\pi } = \Delta _0 \tilde{\pi } $. 
Then 
the set of all full flags that can be constructed by reordering the
basis vectors in $E$ is 
$${\mathcal A} := \{ \Delta _{\pi } \mid \pi \in \Sym_n \}  
= \{ \Delta _0 \tilde{\pi } \mid \pi \in \Sym_n \} .$$

\begin{definition} \label{apart}
The set ${\mathcal A}$ is called the {\bf standard apartment}. 
\end{definition}

The standard apartment ${\mathcal A}$ has a very nice property that 
it contains a system of representatives of the $B$-orbits on 
${\cF }_f(V)$. 
This follows directly from the Gau\ss-Bruhat decomposition: 

\begin{cor}\label{GB}
For all $\Delta \in {\cF }_f(V) $  there is a unique 
$\pi (\Delta ) =: \pi \in \Sym _n$ such that 
$$\Delta b = \Delta _{\pi } \in {\mathcal A} $$
for some $b\in B$.
\end{cor}

The next lemma expresses the well known fact that any two flags have a 
compatible basis. 

\begin{lemma}\label{compbas} 
For any two $\Delta , \Delta ' \in {\cF }_f(V)$ there is some 
$g\in \GL (V)$ such that 
$\Delta g = \Delta _0 $ and $\Delta ' g = \Delta  _{\pi }  \in {\mathcal A} $
for some $\pi \in \Sym _n$, uniquely determined by $\Delta $ and $\Delta '$. 
In particular any $\GL(V)$-invariant distance function 
$d$ on ${\cF}_f(V)$ satisfies $d(\Delta, \Delta ') = 
d(\Delta _0 ,\Delta _{\pi })$.
\end{lemma} 

\bew
As the action of $\GL(V)$ on ${\cF } _f (V)$ is transitive, there is 
some $h\in \GL(V)$ such that $\Delta h = \Delta _0$. 
By Corollary \ref{GB} there is some $b\in B$ such that 
$(\Delta ' h )  b = \Delta _{\pi }$ (with $\pi = \pi (\Delta ' h) \in \Sym _n$). 
Then $g:=hb$ satisfies 
$\Delta g = \Delta _0 b = \Delta _0$ and $\Delta' g = \Delta _{\pi} $
as desired. 
\eb

As any flag can be refined to be a full flag, Lemma \ref{compbas} 
holds equally for non full flags. 

\begin{corollary} \label{compbasgen}
For any two flags $\Lambda , \Lambda ' \in \cF (V) $ (not necessarily of 
the same type) there is some $g\in \GL(V)$ such that 
$\Lambda g$ and $\Lambda ' g$ are contained in full flags
lying in ${\mathcal A}$.
\end{corollary}

\section{Distance functions on spherical buildings}

\subsection{The $\Sym _n$-valued distance function} 

In this section we want to study $\GL (V)$-invariant distance functions 
on $\cF _T(V)$. 
We have seen in Remark \ref{distconst} that such functions 
are constant on the double cosets.
In particular for the full flags $\cF _f(V)$ they factor through the 
$\Sym _n$-valued distance function:

\begin{definition} %\footnote{TODO: Define this also for non full flags} 
The $\Sym _n$-valued distance function of $\cF _f(V)$ is defined as 
$$d_W(\Delta _0 g , \Delta _0 h) = \pi \in \Sym _n \mbox{ if } hg^{-1} \in B\pi B .$$
\end{definition} 

For two flags $\Delta _{\pi }$ and $\Delta _{\sigma } $ in 
the standard apartment ${\mathcal A}$ we have 
$$d_W(\Delta _{\pi } , \Delta _{\sigma } ) = \sigma \pi ^{-1 } .$$

\begin{remark} 
For $\pi \in \Sym _n$ 
the $\pi $-circle around $\Delta _0$ is defined as
$$C_{\pi }(\Delta _0) := \{ \Delta \in \cF(V) \mid d_W(\Delta _0 ,\Delta ) = {\pi } \} .$$
Then
$$C_{\pi }(\Delta _0) = \{ \Delta _0 \tilde{\pi } b \mid b\in B \} 
=\Delta _{\pi } B .$$
The $\pi$-circle $C_{\pi }(\Delta _0)$ is in bijection with the subgroup $U_{\pi }$ mentioned in the end of Section \ref{flags}, 
i.e. for each $\Delta \in C_{\pi }(\Delta _0)$ there is a unique $u\in U_{\pi }$ such that 
$$\Delta = \Delta _0 {\pi } u .$$
In particular if \emph{$K$} is a finite field with $q$ elements, then 
$C_{\pi }(\Delta _0)$ contains exactly $q^{\ell (\pi )}$ 
elements. 
For each ${\pi }\in \Sym _n$ the intersection of the $\pi $-circle 
around $\Delta _0$ with the standard apartment ${\mathcal A} $ defined 
in Definition \ref{apart} is 
$$C_{\pi }(\Delta _0) \cap {\mathcal A} = \{ \Delta _{\pi } \} .$$
\end{remark}

%So the $w$-circle is a regular orbit under the subgroup $U_w$ and 
 %each  $w$-circle intersects the standard apartment ${\mathcal A}$ 
%in a unique element.

\subsection{The Grassmann distance of flags} 

In this section we define a new distance on the set of all 
flags of a given type in $V$, which we call the 
{\bf Grassmann distance of flags}, because it is a direct generalization
of the Grassmann distance on ${\cG}_{k} (V)$ 
the set of subspaces of $V$ of 
dimension $k$. 

\begin{defn}\label{Grassdist}
Let $\Lambda =\{ W_1,\ldots , W_{m} \}$ and 
$\Lambda '  =\{ W'_1,\ldots , W'_{m} \}$  be two  flags in 
$V$ of the same type $T=\{d_1,\ldots , d_m\}$ with
$d_i = \dim(W_i) = \dim (W'_i) \in \{ 1,\ldots, n-1 \}$ for all 
$i$. 
Then the Grassmann distance is defined as
$$\E(\Lambda, \Lambda ') := \sum _{i=1}^{m}
(d_i-\dim (W_i\cap W'_i) ) .$$
\end{defn}

\begin{theorem}\label{GrassdistTheorem} 
For any type $T$, 
the Grassmann distance $\E$ is a $\GL(V)$-invariant distance 
function on the set $\cF _T(V)$ of all flags of type $T$.
\end{theorem}

\bew
Let $\Lambda =\{ W_1,\ldots , W_{m} \}$,
$\Lambda '  =\{ W'_1,\ldots , W'_{m} \}$ 
and 
$\Lambda ''  =\{ W''_1,\ldots , W''_{m} \}$ 
  be flags in
$V$ of the same type $\{ d_1,\ldots , d_m\}$ with
$d_i = \dim(W_i) = \dim (W'_i) = \dim(W''_i)  \in \{ 1,\ldots, n-1 \}$ for all
$i$.
\\
We clearly have that $\E(\Lambda , \Lambda ') = 0 $ if and only 
if $\Lambda = \Lambda' $. 
Also the symmetry $\E(\Lambda ,\Lambda') = \E(\Lambda' , \Lambda )$ is clear. 
The triangle inequality 
$$\E(\Lambda , \Lambda '') \leq \E(\Lambda , \Lambda ') + \E(\Lambda' , \Lambda '')$$
follows from the well known triangle inequality of the
Grassmann distance on subspaces: 
For all $1\leq i \leq m$ we have 
$$ d_i-\dim (W_i\cap W''_i)  \leq 
(d_i-\dim (W_i\cap W'_i)) + (d_i-\dim (W'_i\cap W''_i))  $$
so this also holds for the sum. 
That the function $\E$ is $\GL(V)$-invariant follows directly from the definition.
\eb

For two subspaces $W_i,W'_i $ of dimension $i$ we have 
$$\dim (W_i\cap W'_i) + \dim (W_i + W'_i ) = \dim (W_i ) + \dim (W'_i) = 2i .$$
In particular $\dim (W_i \cap W'_i) \geq 2i -n $ so we have

\begin{remark} (cf. Remark \ref{scoopdist}) 
For two full flags $\Delta, \Delta '$ we have that 
$$\E (\Delta ,\Delta ') \leq \left\{ \begin{array}{ll} (n/2)^2 & n \mbox{ even } \\
(n-1)(n+1)/4 & n \mbox{ odd } \end{array} \right. $$
\end{remark}

The Gau\ss -Bruhat decomposition shows that every
$\GL (V)$-invariant distance on the set of all full flags in $V$ 
factors through $d_W$.
This also holds for the Grassmann distance $\E$,
where $d_W$ is composed with the depth function  $\depth$
from Definition \ref{depth}.

\begin{corollary} %\footnote{Is this also true for non full flags ?} 
For any pair $\Delta , \Delta ' \in \cF _f(V)$ of full flags in 
$V$, we have 
$$\E(\Delta, \Delta ' ) = \depth(d_W(\Delta, \Delta ')) .$$
\end{corollary} 

\bew
By Lemma \ref{compbas} 
it is enough to consider the standard apartment ${\mathcal A}$ 
and to show that for all $\pi \in \Sym _n$ 
$$\E(\Delta _0,\Delta _{\pi } ) = \depth(\pi ) .$$
So let $V_i= \langle e_1,\ldots ,e_i \rangle $ and 
$V'_i = \langle e_{\pi(1)} ,\ldots , e_{\pi(i)} \rangle $
for all $1\leq i \leq n-1$.
Then 
$$V_i\cap V'_i  = 
 \langle e_j \mid j \leq i \mbox{ and }  \pi ^{-1} (j) \leq i \rangle  
 = \langle e_{\pi(k)} \mid \pi(k) \leq i \mbox{ and }  k \leq i \rangle $$
in particular
$$ i -\dim (V_i\cap V'_i) = | \{ k\in \{ 1,\ldots , i \} \mid \pi (k) > i \} | .$$
\eb

\begin{corollary}\label{esym}
As $d_W(\Delta',\Delta ) = d_W(\Delta , \Delta ') ^{-1}$ and
the function $\E $ is symmetric we obtain that 
$\depth(\pi ) = \depth(\pi ^{-1} )$ for all $\pi \in \Sym _n$.
\end{corollary}

\subsection{The gallery distance} 

In the theory of spherical buildings, the most commonly used 
distance function is the 
the gallery distance.
This section compares the Grassmann distance
to the gallery distance.

\begin{definition}\label{galery}
Two full flags $\Delta $ and $\Delta '$ are said to have gallery distance 1
$$d_G(\Delta,\Delta ') = 1 $$
if and only if their intersection $\Delta \cap \Delta $ has cardinality $n-2$,
$$\Delta  \cap \Delta ' = \Delta  \setminus \{W_k \} = \Delta ' \setminus 
\{ W'_{k } \} $$
for some $W_k \in \Delta $, $W'_{k}\in \Delta '$.
A {\bf gallery} is a sequence $\cG = (\Delta _1,\Delta _2,\ldots , \Delta _m)$
of full flags $\Delta _i$ such that $d(\Delta _i,\Delta _{i+1}) = 1$ for
all $1\leq i < m$.
The {\bf length} of the gallery $\cG $ is $m-1$.
It is well known (and follows from elementary linear algebra)
that any two flags $\Delta $ and $\Delta '$ can be joined
by some gallery $\cG = (\Delta, \Delta_1,\ldots , \Delta_{m -1 } , \Delta ')$.
Then the {\bf gallery distance} $d_G(\Delta,\Delta ')$ is the
minimal length $m $ of such a gallery.
\end{definition}

\begin{theorem}(\cite[Section 4.8]{Abramenko}) 
For all $\Delta \in {\cF }(V)$ we have
$d_G(\Delta _0 , \Delta ) = \ell (\pi (\Delta )) $. 
In particular if $\Delta ,\Delta '\in \cF _f(V) $  then 
$$d_G(\Delta , \Delta ')  = \ell (d_W(\Delta , \Delta ')).$$
\end{theorem}

From Theorem \ref{lesym} we now immediately obtain the 
following Corollary.

\begin{corollary}\label{dE}
If $\Delta \neq \Delta '  \in {\cF }_f(V)$ are full flags in $V$, then
$$ 2\E(\Delta, \Delta ') > d_G(\Delta, \Delta') \geq \E(\Delta , \Delta ') .$$
\end{corollary}

\section{The channel model}

Throughout this section we will work with a fixed type 
 $T = \{ d_1,d_2,\ldots, d_{m}\}$ with
$0<d_1<\ldots < d_m < n$, and put 
$k_i := d_i - d_{i-1}$, $i=2,\ldots ,m$, $k_1:=d_1$.

We will model our network as a finite 
directed, acyclic multigraph with a single source and possibly multiple receivers. 
Every edge gets a capacity of $1$, but we allow multiple edges between nodes to model different capacities. 
The source and the receivers agree on a set  $\cC\subset \cF_T(V)$
of flags of type $T$, 
the error correcting code. 
Information is encoded as a flag $\Lambda  \in \cC$. \\
Assume now that the source has a flag $\Lambda = 
\{ V_1 , V_2 , \ldots , V_m \} \in \cC $ 
with $d_i = \dim (V_i)$. Fixing a basis of $V$ and therewith identifying 
$V$ with the space of rows, $K^n$, we may think 
of $\Lambda $ as 
 a sequence of row vectors
$x_1,x_2, \ldots x_{d_m}\in K^n$ such that $x_1,x_2,\ldots, x_{d_i}$ form a basis of
$V_i$. 
For $1\leq j\leq m$ let 
$X_j \in K^{d_j \times n}$ be the matrix whose $i$-th row is $x_i$. 
Of course $X_j$ is a submatrix of $X_{j+1}$ and so all 
the information is contained in the matrix $X_m$.
\\
At every time step $1 \leq i \leq m$ and for every
outgoing edge the source chooses  random coefficients $y\in K^{1\times d_i}$ 
and sends  $y \cdot X_i \in V$  through that edge. \\
Furthermore every intermediate node forms a random linear combination of everything received up to this point for every edge. \\
So at time $i$ the receiver receives many (say $a_i$)  random linear 
combinations of the rows $x_1,\ldots , x_{d_i} $, i.e. $Z_i= Y_i \cdot X_i$ with
$Y_i\in K^{a_i\times d_i }$.
The receiver defines spaces 
$$W_i := \langle  \text{ rows of } Z_j \mid 1\leq j\leq i \rangle. $$
Then by definition 
 $W_i \leq W_{i+1}$ for all $i$. 
Put $\Gamma := (W_1 , W_2 , \ldots , W_m)$.

 \begin{remark}
	 If $Z_i = Y_i \cdot X_i$ and 
	 the rank  of the matrix  formed by the 
	 last $k_i$ columns of  $Y_i$ equals $k_i$ for all $i$,
	 then $W_i = V_i$ for all $i$ and 
	 $\Lambda = \{ W_1,\ldots , W_m \}$.
This is the case if there are no errors or erasures in the transmission. 
	 Note that a necessary condition is that each $Y_i$ has 
	 at least $k_i$ rows, so all the $k_i$ need to be at most the
	 capacity of the network.
 \end{remark}

Note however that due to erasures or errors the receiver 
gets some matrix $\tilde{Z}_i = Y_i X_i + E_i $ 
where the rank of $Y_i$ is smaller than $d_i$ (due to erasures) 
and $E_i \neq 0$ (due to errors). 
We then might have that $W_i\neq V_i$, and $\Gamma $ might not even
be a flag of length $m$, but only a stuttering flag in the sense of
the following definition.

\begin{definition}
	\begin{itemize}
		\item[(a)] 
	A {\em stuttering flag} of length $m$ is a sequence 
$\Gamma := (W_1 , W_2 , \ldots , W_m)$ of subspaces of $V$ such that 
$W_1\leq W_2 \leq \ldots \leq W_m \leq V$.
\item[(b)] Let $\Lambda =\{V_1,\ldots , V_m\} \in \cC $ be the sent flag and 
	$\Gamma  = (W_1 , W_2 , \ldots , W_m) $ the received stuttering flag.
	In analogy to \cite[Definition 1]{KK} 
 we define 
$$\rho _i = \rho_i(\Lambda , \Gamma ) := \dim (V_i) - \dim (W_i\cap V_i)$$
to be the number of erasures in step $i$ and 
$$f_i = f_i(\Lambda , \Gamma ) := \dim (W_i) - \dim (W_i\cap V_i)$$
the number of errors in step $i$.
\item[(c)] 
	The final error count between  $\Lambda $ and $\Gamma $ is 
$$E(\Lambda,\Gamma ) :=  \sum_{i=1}^m \dim(V_i + W_i) - \dim(V_i \cap W_i).$$
\end{itemize}
\end{definition}

\begin{remark}
	The final error count satisfies
	$E(\Lambda , \Gamma ) = \sum_{i=1}^m (\rho_i(\Lambda ,\Gamma ) + f_i (\Lambda, \Gamma ))$. 
\end{remark}

\bew
This follows from the famous Grassmann identity: \\
$\dim (V_i+W_i) + \dim (V_i \cap W_i) = \dim(V_i) + \dim(W_i) .$
\eb

Note that if $\Lambda $ and $\Gamma $ are both flags of type $T$
then $E(\Lambda , \Gamma ) = 2 \E (\Lambda ,\Gamma)$. 
The error count
also originates from the Grassmannian distance on subspaces and is thus a metric satisfying the triangle inequality. 
Hence in analogy to \cite[Theorem 2]{KK} we get the following corollary.

\begin{corollary}
Let $\cC$ be a set of flags of type $T$ and 
$$d(\cC) := \min \{\E(\Lambda', \Lambda) \mid \Lambda ' \neq \Lambda \in \cC\}$$
be the minimum distance of $\cC $.
Using the code  $\cC $ for transmission through the network we can correct all errors as long as the error count satisfies
$E(\Lambda , \Gamma) < d(\cC) ,$ meaning that in this case $\Lambda $ 
is the unique element of $\cC $ such that $E(\Lambda,\Gamma )$ is minimal.
\end{corollary}

\bew
Let $e := E(\Lambda, \Gamma)$. For another flag $\Lambda \neq \Delta \in \cC$ set $f := E(\Delta, \Gamma).$ Then the triangle inequality gives us
$$E(\Lambda, \Delta) \leq E(\Lambda, \Gamma) + E(\Gamma, \Delta) = e + f.$$
On the other hand $\Lambda$ and $\Delta$ are elements of $\cC$ and thus we can use the observation from above to get
$$d(\cC) \leq \E(\Lambda, \Delta) = \frac{E(\Lambda, \Delta)}{2}.$$
Putting these together we get $2d(\cC) \leq e+f$. But as we assumed that $e < d(\cC)$ we thus have $f > d(\cC) > e$, hence $\Lambda$ is 
indeed the unique element of $\cC$ having minimal distance to $\Gamma$.
\eb

\subsection{Error correcting codes} 

For good error correcting codes, as in the classical situation, 
$|\cC |$ and $d(\cC )$ both should be large. 
%But we also face the problem of efficient decoding.

One idea is to construct $\cC $ as an orbit 
$\Delta _T S$ of some subgroup
$S\leq \GL(V) $ with $S\cap P_T = \{ 1 \}$. 
Then, using Remark \ref{distconst} we can compute the
minimum distance on $\cC= \Delta _T S $ as follows:

\begin{remark} \label{distmat1} 
For $g,h\in \GL(V)$ we have 
$$\E(\Delta _T g , \Delta _T h ) =\E(\Delta _T (gh^{-1}), \Delta _T) =: \overline{\E}_T(gh^{-1}) .$$
In particular if $S\leq \GL(V) $ with $S\cap P_T = \{ 1\}$, 
then 
$$d(\Delta _T S) = d_T(S) := \min \{ \overline{\E}_T(g)\mid 1\neq g\in S \} .$$
As usually we abbreviate $\overline{\E}_{\{ 1,\ldots , n-1 \} } $ 
by $\overline{\E}_f$.
\end{remark}

\begin{lemma}\label{distmat}
Let $T = \{ d_1,\ldots , d_m \}$, 
$k_i:=d_i-d_{i-1}$ ($2\leq i\leq m$), $k_1:=1$, $k_{m+1} :=n-d_m$,  and
$$g = \left(\begin{array}{ccccc} 
I_{k_1} & A_{12} & \ldots & \ldots & A_{1m} \\ 
0 & I_{k_2} & A_{23} & \ldots & A_{2m} \\
\vdots & \ddots & \ddots & \ddots & \vdots  \\ 
0 & \ldots & 0 & I_{k_{m}} & A_{mm}  \\ 
0 & \ldots & \ldots & 0 & I_{k_{m+1}} \end{array} \right) $$
where $A_{ij} \in K^{k_i \times k_j }$.
For $i=1,\ldots , m$ put 
$$g_i := 
\left(\begin{array}{ccc} 
 A_{1,i+1} & \ldots & A_{1m} \\ 
\vdots & \vdots & \vdots  \\ 
 A_{i,i+1} & \ldots & A_{im}  
\end{array} \right) $$
the upper right $d_i \times (n-d_i)$ submatrix of $g$ and $r_i = \rk(g_i)$.
Then $\overline{\E}_T(g) = \sum _{i=1}^m r_i .$
\end{lemma}

\bew
Let $\Delta_T = (V_1,V_2,\ldots, V_m)$ and $\Lambda = \Delta_T g = (W_1,W_2,\ldots, W_m)$ where $\dim(V_i) = \dim(W_i) = d_i$ for all $i$.
We claim that $r_i = d_i - \dim(V_i\cap W_i)$, again for all $i$. The proof of the lemma follows directly from that claim by writing a sum on both sides. \\
To prove the claim fix one $i$ and consider the matrix
$$M = \begin{pmatrix} \begin{matrix}
        I_{k_1} & A_{12} & \ldots & \ldots \\ 
        0 & I_{k_2} & A_{23} & \ldots  \\
        \vdots & \ddots & \ddots & \ddots   \\ 
0 & \ldots & 0 & I_{k_{i}}  \end{matrix}  & 
\begin{matrix} A_{1,i+1}  & \ldots & A_{1m} \\
               A_{2,i+1} & \ldots & A_{2m} \\
               \vdots & \vdots & \vdots \\
               A_{i,i+1} &  \ldots & A_{im} \end{matrix} \\
\Scale[2]{I_{d_i}} & \Scale[2]{0}\end{pmatrix}. $$
Then the row space of $M$ equals $V_i+W_i$ and thus the rank of $M$ equals
$$\rk(M) = \dim(V_i+W_i) = 2d_i-\dim(V_i\cap W_i).$$
To compute the rank of $M$ we use Gau\ss{}ian elemination. As we have a big identity matrix on the bottom we can use that to reduce $M$ to the matrix
$$\begin{pmatrix}0 & g_i \\ I_{d_i} & 0\end{pmatrix}.$$
Now we see $\rk(M) = r_i+d_i$ and this gives us
$$r_i+d_i = \rk(M) = 2d_i-\dim(V_i\cap W_i) \,\,\,\, \Rightarrow\,\,\,\, r_i = d_i-\dim(V_i\cap W_i).$$
\eb

In fact we retrieve the idea of \cite{SKK} here:

\begin{example}\label{codes0}
Let $T:= \{ k \} $, so $\cF _T(V)$ correspond to the
Grassmannian $\cG _k(V)$. 
Choose some
subspace $C\leq K^{k\times n-k} $ 
and put
$$U_C := \{ u(c) := \left( \begin{array}{cc} I_k & c \\ 0 & I_{n-k} \end{array} \right)
\mid c\in C \} .$$
Then $U_C$ is a subgroup of $\GL (V)$ isomorphic to the additive group of $C$.
In particular for $c,c' \in C$ we compute 
$u(c) u(c')^{-1} = u(c-c') $. 
So Lemma \ref{distmat} is a direct generalization of 
\cite[Proposition 4]{SKK} that the Grassmann distance on the subspaces 
with basis matrix $(I_k|c)$ is the rank metric on $K^{k\times n-k }$.
\end{example} 

To compare such commonly used subspace codes with 
our new flag codes assume for convenience that 
$n=4m$ and $C  \leq K^{2m\times 2m}$ is an
MRD code of dimension $2m$
with rank metric distance
$$d(C) := \min \{ \rk (c) \mid 0\neq c\in C \} = 2m $$
 (see for instance \cite[Section C]{SKK}).
Then 
$$\dim (C) = 2m = \dim (U_C) \mbox{ and } 
d(C) = d_{\{2m \} } (U_C) = 2m .$$
Using flags of type $T=  \{ m,2m,3m \} $ we can improve on the
dimension of the flag code (we get dimension  $3m$) keeping the 
minimum distance to be  $2m$:

\begin{prop}\label{codes1}
Assume that $n=4m$ and put $T:= \{ m,2m,3m \} $.
Given two MRD codes $C_m$ and $C_{2m}$ 
with $$C_i \leq K^{i\times i}, d(C_i) = i, \dim (C_i ) = i $$
we
put
$$\cC (C_{m},C_{2m}):= \{ \Delta _{T} u(x,y)  \mid x\in C_m, y\in C_{2m} \} \subseteq \cF _T(V)$$
where
$$ u(x,y) := \left(\begin{array}{cc} 
\begin{array}{cc} I_m & x  \\ 0 & I_m \end{array} 
& y  \\
0 & 
\begin{array}{cc} I_m & x  \\ 0 & I_m \end{array}  \end{array} 
\right) \in \GL_{4m} (K) $$ 
Then 
$\dim (\cC(C_{m},C_{2m})) = 3m $ and 
$d(\cC(C_{m},C_{2m})) = 2m .$
\end{prop} 

For the proof we need the following elementary fact
about multiplication of block triangular matrices.

\begin{lemma}\label{multmat}
Let $A \in \GL_m(K)$, $B,D\in K^{m\times m} $.
Then 
$$\left( \begin{array}{cc} 
A & B \\ 0 & A \end{array} \right) ^{-1 } = 
\left( \begin{array}{cc} 
A^{-1} & -A^{-1}BA^{-1} \\ 0 & A^{-1} \end{array} \right) $$ 
and 
$$\left( \begin{array}{cc} 
A & D \\ 0 & A \end{array} \right) 
 \left( \begin{array}{cc} 
A & B \\ 0 & A \end{array} \right) ^{-1 }= 
\left( \begin{array}{cc} 
I_m & (D-B)A^{-1} \\ 0 & I_m \end{array} \right) .$$ 
\end{lemma} 

\bew (of Proposition \ref{codes1})
For $0\neq x\in C_m$, the rank of $x\in K^{m\times m}$ is $m$ and hence 
$\overline{\E}_T(u(x,y)) \geq 2m $ for any $y\in K^{2m\times 2m}$ by Lemma \ref{distmat}. 
Now
 $u(x,y)^{-1} = u(-x,y'' )$ for some $y'' \in K^{2m\times 2m} $,
hence $u(x',y') u(x,y)^{-1} = u(x'-x,y''') $ for some $y'''$ so 
by Remark \ref{distmat1}. 
$$\E(\Delta _T u(x,y) , \Delta _T u(x',y') ) = \overline{\E}_T(u(x'-x,y''')) \geq 2m \mbox{ if } x\neq x'.$$ 
Assume that $x=x'$  then by Lemma \ref{multmat} 
$ u(x,y') u(x,y)^{-1} = u(0,y'') $
with 
$$y'' = (y'-y) \left( \begin{array}{cc}
1 & -x \\ 0 & 1 \end{array} \right)
$$
in particular the rank of $y'' $ is the same as the one of 
$y-y'$.
If $y\neq y'$ then this is a non zero element of 
$C_{2m}$ so it has rank $2m$. 
Using Remark \ref{distmat1} we again 
find $\E (\Delta _T u(x,y) , \Delta _T u(x,y') ) = 2m $.
\eb

\subsection{Checkerboard codes}

We now want to iterate the idea from Proposition 
\ref{codes1}.
Assume that we have a sequence of MRD codes 
$C_i \leq K^{2^i\times 2^i} $ 
$(i=0,\ldots , t)$ such that 
$$\dim _K(C_i ) = 2^i,\ \ \rk (c) = 2^i \mbox{ for all } 0\neq c\in C_i .$$

\begin{definition} \label{codes2} 
For $x_i \in C_i $ ($0\leq i \leq t$) we define 
$$u(x_0,x_1, \ldots ,x_t) \in \GL_{2^{t+1}} (K) $$ 
recursively as 
$$u(x_0) = \left(\begin{array}{cc} 1 & x_0 \\ 0 & 1 \end{array}\right) , \ 
u(x_0,\ldots , x_t) = \left( \begin{array}{cc} 
u(x_0,\ldots , x_{t-1}) & x_t \\ 0 & 
u(x_0,\ldots , x_{t-1}) \end{array}\right) .$$
Then the 
{\bf checkerboard code}  associated to the MRD codes $C_i$ is
$$\cC (C_0,C_1,\ldots , C_t) = 
\{ \Delta _0 u(x_0,x_1, \ldots ,x_t) \mid x_i \in C_i \mbox{ for } 
0\leq i \leq t \} \subset {\cF }_f (V).$$
\end{definition} 

Note that the dimension of $\cC (C_0,C_1,\ldots , C_t) $ is 
$  \sum _{i=0}^t 2^i = 2^{t+1} -1 .$

\begin{prop}
Let $\cC := 
\cC (C_0,C_1,\ldots , C_t) $.  \\
Then $\dim  (\cC) = 2^{t+1}-1 $ and $d (\cC ) = 2^{t} $.
\end{prop} 

\bew
We show by induction on $t$ that 
$d (\cC (C_0,C_1,\ldots , C_t)) \geq 2^{t} $.
 For $t=0$ there is nothing to show. 
So let 
$$
g:=\left( \begin{array}{cc} 
A & B \\ 0 & A \end{array} \right), 
h:=\left( \begin{array}{cc} 
A' & B' \\ 0 & A' \end{array} \right) \in \GL _{2^{t+1}} (K)  $$
where $A,A' \in \{ u(x_0,\ldots , x_{t-1}) \mid x_i \in C_i \}$,
$B,B' \in C_t$. 
By Remark \ref{distmat1} we need to show that $\overline{\E} _f (hg^{-1}) \geq 2^t$,
if $g\neq h$.  By Lemma \ref{multmat}  
$$hg^{-1} = \left( \begin{array}{cc} 
A'A^{-1} & B'' \\ 0 & A'A^{-1 } \end{array} \right) $$
with $B'' = (B'-B) A^{-1} $ if $A=A'$. 
If $A\neq A'$ then 
$$\overline{\E}_f(hg^{-1} ) \geq 2 \overline{\E}_f (A'A^{-1}) \geq 2 \cdot 2^{t-1}  = 2^t $$
by induction. 
If $A=A' $ then $B'\neq B \in C_{t}$ (because $g\neq h$) and hence 
$\rk (B'-B) = 2^t$ so $\rk (B'') = 2^t$ and 
$\overline{\E}_f(hg^{-1} ) \geq 2^t $ by Lemma \ref{distmat}.
Note that 
$\overline{\E}_f(u(x_0,0,\ldots ,0) ) = 2^t $ by Lemma \ref{distmat} so
$d (\cC (C_0,C_1,\ldots , C_t)) \leq 2^{t} $ and hence 
we get the equality as claimed. 
\eb

\subsection{Derived subgroup codes}

Take $D$ to be the subgroup of all upper uni-triangular matrices in $\GL_n(K)$:
$$D = \left\{ \left( \begin{array}{cccc} 
1 & * & * & * \\
0 & 1 & * & * \\
0 & 0 & \ddots & \vdots \\
0 & 0 & 0 & 1\end{array} \right) \mid * \in K \right\} \leq \GL_n(K).$$
Then the derived subgroups of $D$ are of the form
$$D^{(k)} = \{g \in D \mid g_{ij} = 0 \text{ for } 0 < j-i \leq k \},$$
thus having exactly $k$ secondary diagonals filled with zeros. 

\begin{prop}\label{derived} 
The code $\cC (n,k) := \Delta_0 D^{(k)}$ consists of fine flags and 
has parameters
$$d(\cC (n,k)) = k+1,\ \dim(\cC (n,k)) = \frac{(n-k)(n-k-1)}{2} .$$
\end{prop}

\bew
To compute the dimension of $\cC (n,k) $ we just count the 
number of free parameters to be 
$$\sum _{j=1}^{n-k-1} j = \frac{(n-k)(n-k-1)}{2} .$$
For computing  the minimal distance we use the 
fact that 
$D^{(k)}$ is a group. So it suffices to compute $\overline{\E}_f(g)$ for $1\neq g \in D^{(k)}$.
Then $g$ has at least one non-zero entry at a position $(i,j)$ with $j \geq i+k+1$.
Using Lemma \ref{distmat} this gives us $j-i \geq k+1$ matrices with rank at least one, hence $\overline{\E}_f(g) \geq k+1$. \\
On the other hand taking a $g \in D^{(k)}$ that only
 has one non-zero entry at position $(i,i+k+1)$ for some $i$ yields a matrix with $\overline{\E}_f(g) = k+1$.
\eb

\begin{rem}
The code $\cC(n,k)$ allows for decoding erasures 
using only Gau\ss{}ian elimination. If we receive a stuttering flag
$$\Lambda = (U_1,\ldots, U_{n-1})$$
we can uniquely recompute the corresponding matrix $g \in D^{(k)}$ as long as the longest subchain $U_i,U_{i+1},\ldots,$ of subspaces
such that  $\dim(U_{i+j}) < i+j$ has length at most $k$. 
\end{rem}

\bew
Let $g_j$ be the submatrix of $g$ consisting of the first $j \leq n$ rows. Then due to the zeros
on the secondary diagonals the last $k+1$ rows of $g_j$ 
are not changed when computing the reduced row echelon form of $g_j$. 
If we receive a subspace $U_j$ with $\dim(U_j) = j$ we can hence compute a reduced row echelon form of a generator matrix
of $U_j$ and by the uniqueness of that form we get the $i$-th row of $g$ for all $j-k \leq i \leq j$.
Thus we can recompute $g$ as long as we have that at least every $k-$th space in $\Lambda$ has the correct dimension.
\eb


{\small
\begin{thebibliography}{99}
\bibitem{Abramenko} P. Abramenko, K.S. Brown, {\em Buildings} 
Theory and Applications, Springer Graduate Texts in Mathematics 248 (2008)
\bibitem{Ahlswede} 
	R. Ahlswede, N. Cai, S.-Y.R.  Li, R.W. Yeung, 
	Network information flow. 
	IEEE Trans. Inform. Theory 46 (2000) 1204--1216.
\bibitem{Humphreys} J.E. Humphreys, Reflection Groups and Coxeter Groups, 
Cambridge studies in advanced mathematics 29 (1990) 
\bibitem{Jones} 
A.R. Jones,  A combinatorial approach to the double cosets of the symmetric group
 with respect to Young subgroups.
 European J. Combin.  17  (1996) 647--655.
\bibitem{KK} R. K\"otter, F.R. Kschischang,  
Coding for errors and erasures in random network coding.
 IEEE Trans. Inform. Theory  54  (2008) 3579--3591.
\bibitem{SKK} 
D. Silva,
 F.R. Kschischang,  
R. K\"otter,
 A rank-metric approach to error control in random network coding.
 IEEE Trans. Inform. Theory  54  (2008) 3951--3967.
\bibitem{Taylor} D.E. Taylor, The geometry of the classical groups. Heldermann Verlag Berlin (1992)
\bibitem{OEIS} 
N.J.A. Sloane, editor, The On-Line Encyclopedia of Integer Sequences, published electronically at https://oeis.org (31.05.2016) 
\bibitem{A-ITW} M.A. Vazquez-Castro. 
A geometric approach to dynamic network coding. 
 IEEE Information Theory Workshop, 2015. 
 \bibitem{Petersen} T.K. Petersen, B.E. Tenner, The depth of a permutation, Journal of Combinatorics 6, (2015) 145--178
\end{thebibliography}
}
\end{document}